\documentclass[nonacm,sigplan]{acmart}

\usepackage[]{hyperref}
\usepackage{multirow}
\usepackage{colortbl}
\usepackage{tabularx}
\usepackage{algorithm}
\usepackage{algorithmic}

\usepackage{graphicx}
\usepackage{subcaption}
\usepackage{tabularx}
\usepackage{float}
\usepackage{xcolor} 

\definecolor{Gray}{gray}{0.95}

\newcommand{\gr}{\rowcolor[gray]{.95}}

\begin{document}

\title{\bf COMET: Towards Practical W4A4KV4 LLMs Serving }


\author{Lian Liu}
\affiliation{%
    \institution{Institute of Computing Technology, CAS, University of Chinese Academy of Sciences}
  \city{Beijing}
  \country{China}
  \postcode{100190}
}
\email{liulian21s@ict.ac.cn}

\author{Haimeng Ren}
\affiliation{%
  \institution{ShanghaiTech University}
  \city{Shanghai}
  \country{China}}
\email{renhm2022@shanghaitech.edu.cn}

\author{Long Cheng}
\affiliation{%
  \institution{North China Electric Power University}
  \city{Beijing}
  \country{China}
}
\email{lcheng@ncepu.edu.cn}

\author{Zhaohui Xu}
\affiliation{%
  \institution{ShanghaiTech University}
  \city{Shanghai}
  \country{China}}
\email{xuzhh12022@shanghaitech.edu.cn}

\author{Yudong Pan}
\affiliation{%
  \institution{Institute of Computing Technology, CAS, University of Chinese Academy of Sciences}
  \city{Beijing}
  \country{China}}
\email{panyudong23s@ict.ac.cn}

\author{Mengdi Wang}
\affiliation{%
  \institution{State Key Lab of Processors, Institute of Computing Technology,  Chinese Academy of Sciences}
  \city{Beijing}
  \country{China}}
\email{wangmengdi@ict.ac.cn}

\author{Xiaowei Li}
\affiliation{%
  \institution{State Key Lab of Processors, Institute of Computing Technology,  Chinese Academy of Sciences}
  \city{Beijing}
  \country{China}}
\email{lxw@ict.ac.cn}

\author{Yinhe Han}
\affiliation{%
  \institution{State Key Lab of Processors, Institute of Computing Technology,  Chinese Academy of Sciences}
  \city{Beijing}
  \country{China}}
\email{yinhes@ict.ac.cn}

\author{Ying Wang}
\authornote{Corresponding author.\\[2pt] \textit{PS: This paper was submitted to ASPLOS in June 2024.}}

\affiliation{%
  \institution{State Key Lab of Processors, Institute of Computing Technology,  Chinese Academy of Sciences}
  \city{Beijing}
  \country{China}}
\email{wangying2009@ict.ac.cn}


\begin{abstract}
Quantization is a widely-used compression technology to reduce the overhead of serving large language models (LLMs) on terminal devices and in cloud data centers. However, prevalent quantization methods, such as 8-bit weight-activation or 4-bit weight-only quantization, achieve limited performance improvements due to poor support for low-precision (e.g., 4-bit) activation. This work, for the first time, realizes practical W4A4KV4 serving for LLMs, fully utilizing the INT4 tensor cores on modern GPUs and reducing the memory bottleneck caused by the KV cache. Specifically, we propose a novel fine-grained mixed-precision quantization algorithm (FMPQ) that compresses most activations into 4-bit with negligible accuracy loss. To support mixed-precision matrix multiplication for W4A4 and W4A8, we develop a highly optimized W4Ax kernel. Our approach introduces a novel mixed-precision data layout to facilitate access and fast dequantization for activation and weight tensors, utilizing the GPU's software pipeline to hide the overhead of data loading and conversion. Additionally, we propose fine-grained streaming multiprocessor (SM) scheduling to achieve load balance across different SMs. We integrate the optimized W4Ax kernel into our inference framework, COMET, and provide efficient management to support popular LLMs such as LLaMA-3-70B. Extensive evaluations demonstrate that, when running LLaMA family models on a single A100-80G-SMX4, COMET achieves a kernel-level speedup of \textbf{$2.88\times$} over cuBLAS and a \textbf{$2.02 \times$} throughput improvement compared to TensorRT-LLM from an end-to-end framework perspective.

\end{abstract}

\maketitle 
\pagestyle{plain} 

\section{Introduction}
Large language models (LLMs) have demonstrated excellent performance across various benchmarks~\cite{brown2020language, dettmers2022llm, touvron2023llama, workshop2022bloom, zhang2022opt, openai2023gpt4}. However, as LLMs advance and models with hundreds of billions of parameters emerge, their substantial sizes present significant challenges for inference systems. Specifically, large models require extensive memory, while most LLM-based systems perform inference on a single GPU with limited memory. Moreover, LLM inference incurs high serving costs, often calculated per token, and processing long token sequences further increases these costs.

Model quantization is an efficient way to reduce the memory footprint and serving costs for LLM inference, with weight-only quantization being a typical method in recent years~\cite{frantar2022gptq, lin2023awq}. However, the latest studies~\cite{zhao2024atom, lin2024qserve} report that weight-only quantization achieves limited performance improvements on modern GPUs, particularly when processing large-batch and long token sequences. The main reasons are: (1) weight-only quantization requires low-bit parameters to be dequantized to align with high-precision activations before being processed by the GPU tensor cores, leading to a waste of computational resources. For example, existing W4A16 quantization methods~\cite{frantar2022gptq, lin2023awq, shao2023omniquant} must restore the quantized 4-bit weights to 16-bit and process them together with activations in the FP16 tensor cores, which is inefficient for modern GPUs like A100 that are optimized for higher-throughput 4-bit operations; and (2) in applications involving large-batch processing and long token sequences~\cite{hooper2024kvquant, yue2024wkvquant}, the Key and Value activation caching (KV cache) becomes the major bottleneck rather than the weight parameters. Although methods like SmoothQuant~\cite{xiao2023smoothquant} simultaneously quantize both activations and weights, they employ a conservative scaling strategy that restricts activations and weights to INT8 format, still facing the issues mentioned above.

To address the above issues, achieving lower-precision (e.g., 4-bit) activations without compromising accuracy is crucial. This would fully utilize low-bit tensor cores in modern GPUs, delivering higher throughput. Additionally, low-precision quantization for the KV cache, which consumes significant memory in transformers, is needed. This would alleviate the memory bottleneck, enable larger inference batch sizes, and efficiently exploit the batch-level parallelism in advanced GPUs. These requirements, along with the concentrated distribution of outliers in activations~\cite{dettmers2022llm, sun2023simple}, motivate us to design a novel fine-grained mixed-precision quantization algorithm (FMPQ) for activations. Specifically, FMPQ quantizes most activations to 4-bit and others to 8-bit\footnote{Throughout this paper, \textit{mixed-precision} refers to a combination of W4A4 and W4A8 (i.e., \textbf{W4Ax}), rather than mixing 8-bit or 16-bit activations with lower-precision weights as seen in current works~\cite{lin2023awq, lin2024qserve}.}. To ensure efficient computing, we partition the activation tensor into multiple sub-tensors, each sized to match the GPU's computational granularity. Additionally, we introduce a channel permutation strategy to cluster outliers within the same sub-tensor, thereby reducing the overall quantization precision. 

Since FMPQ requires hardware capable of W4Ax matrix multiplication for the quantized LLM, but existing LLM serving systems~\cite{tensorrt-llm, llama.cpp, rasley2020deepspeed} lack support for direct mixed-precision tensor operations and W4Ax computing, we further design a novel W4Ax kernel and integrated it into our inference framework, COMET. Generally, COMET optimizes mixed-precision LLM computing on GPUs by incorporating data layout design and fast dequantization for mixed-precision weights and activations. It further uses the software pipeline to overlap the overhead of data loading and conversion. Additionally, given that the lower precision tensor cores in modern GPU provide higher throughput (INT4 tensor core has $2\times$ higher throughput than INT8), COMET employs a fine-grained SM scheduling strategy to achieve load balance across different stream multiprocessors (SMs). By integrating the optimized W4Ax kernel and efficient memory management techniques~\cite{kwon2023efficient}, COMET provides practical and efficient LLM serving with the high-performance mixed-precision encoding.

In a nutshell, the contributions of this work can be summarized as follows:

\begin{itemize}
    \item We analyze the distribution of outlier values in LLM activations and introduce a novel FMPQ algorithm that enables 4-bit activations and KV cache without compromising accuracy. To achieve this, the activation tensor is divided and quantized at a granularity that matches the matrix multiplication units on modern GPUs, employing a tiling approach. With negligible accuracy loss, the proposed FMPQ algorithm processes more than 84\% of GEMM computations using W4A4, while W4A8 is used for the remaining computations.
    
    \item We develop a novel highly-optimized W4Ax kernel to support the simultaneous computation of W4A4 and W4A8. The low-precision data points are packed into a high-precision format and directly processed in CUDA cores using an optimized pipeline, effectively hiding the expensive runtime numerical-format conversion overhead. Furthermore, we propose an efficient fine-grained SM scheduling solution during LLM compilation stages. This solution remaps the mixed-precision tensor tiles, to achieve balanced mixed-precision computing across different SMs.

    \item We present COMET, the high-performance mixed-precision LLM inference framework, which integrates our proposed W4Ax kernel and provides efficient memory management for LLM serving. Compared with state-of-the-art (SOTA) frameworks, COMET further improve the performance of LLM serving with mixed-precision quantization. Evaluated on a single A100-80G-SXM4 across various LLM models, COMET demonstrates a $2.02\times$ improvement in end-to-end performance over SOTA baselines. Additionally, we provide an open-source W4Ax kernel with a Python interface and a set of C++ APIs, enabling seamless integration into existing inference systems such as TensorRT-LLM~\cite{tensorrt-llm} and DeepSpeed~\cite{rasley2020deepspeed}.


\end{itemize}

\section{Background \& Motivation}

\subsection{LLM Inference}\label{sec:background-llm-inference}

\begin{figure}
    \centering
    \includegraphics[width=0.98\linewidth]{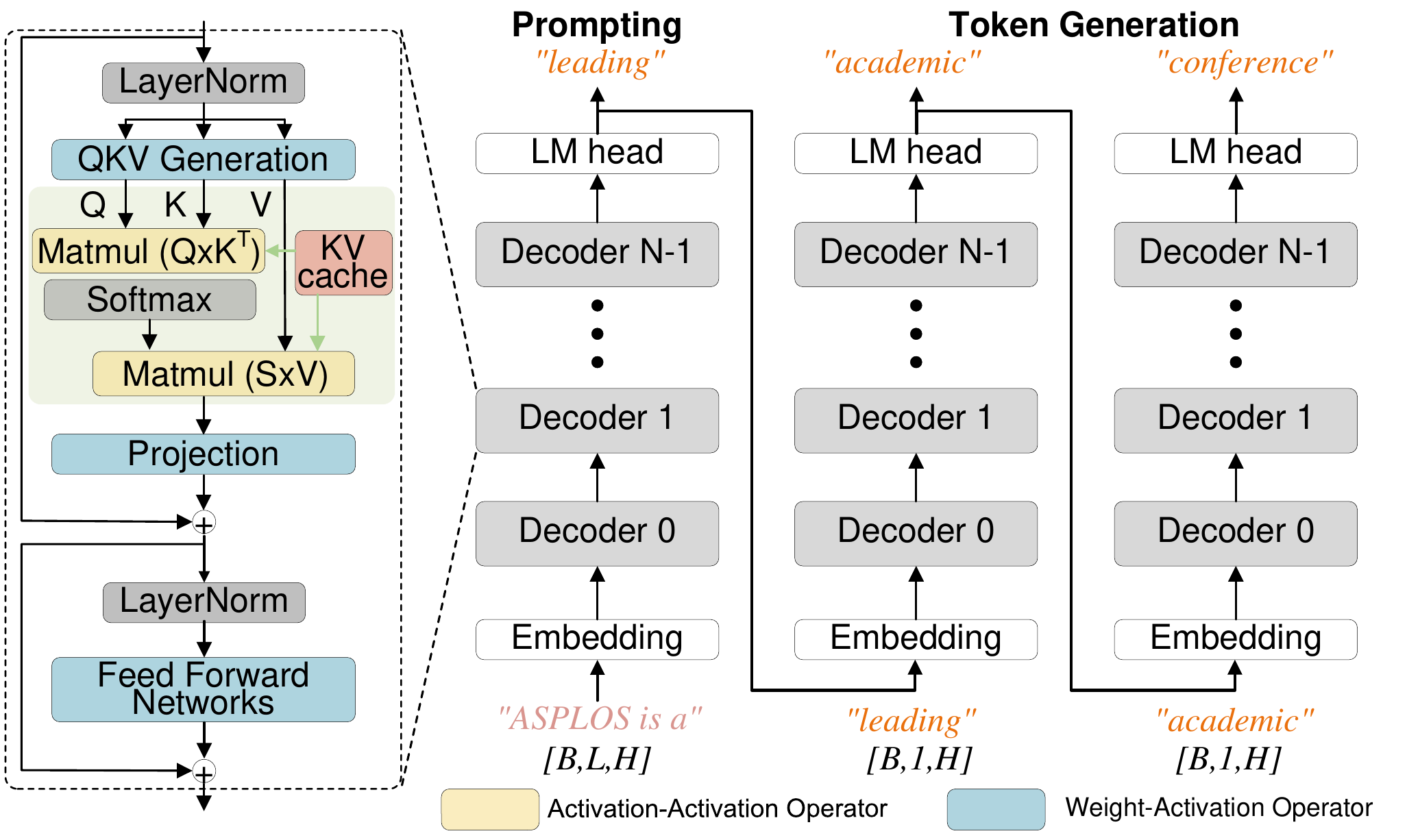}
    \caption{The inference procedure of LLMs. The inference procedure can be divided into two phases: prompting and token generation.}
    \label{fig:llm-execution}
\end{figure}


Figure \ref{fig:llm-execution} illustrates the inference procedure of LLMs, which can be divided into two phases: prompting and token generation. In the prompting phase, the model receives a sentence and processes it in parallel. During the token generation phase, it generates the output sequence auto-regressively, producing one token at a time and using it as input for the subsequent step. Specifically, during prompting, multiple tokens are fed into the model simultaneously and processed in parallel. The input tensor shape during this phase is $[B, L, H]$, where $B$, $L$, and $H$ represent the inference batch size, prompting sequence length, and the hidden dimension of the model, respectively. However, during token generation, the input tensor shape changes to $[B, 1, H]$ since only one token is processed at a time. 

It is important to note that during token generation, the model needs to store intermediate Key and Value activations to avoid redundant computations. This storage is commonly referred to as the KV cache. During inference, the system must preserve not only all weight parameters but also the Key-Value pairs generated during the token generation phase. While the storage space for weights remains constant, the size of the KV cache is directly proportional to the sequence length. Consequently, as sequence length increases, the KV cache gradually becomes the primary storage bottleneck, overtaking weight parameters. For instance, at a sequence length of 128K, the KV cache accounts for 72\% of the total storage cost for LLaMA-7B~\cite{hooper2024kvquant}.

\subsection{LLM Quantization}\label{sec:background-quantization}

Quantization is an effective method for reducing the inference cost of LLMs. Existing quantization methods for LLMs fall into two categories: weight-only and weight-activation quantization~\cite{yue2024wkvquant, zhao2023survey}.

\textbf{Weight-only Quantization. }
Weights are more amenable to quantization compared to activations~\cite{frantar2022gptq, lin2023awq, ashkboos2024quarot}. Consequently, a substantial body of research has concentrated on achieving low-precision weights. Given the high retraining costs associated with LLMs, existing studies mainly focus on post-training quantization (PTQ). Methods such as GPTQ \cite{frantar2022gptq}, QuIP \cite{chee2024quip}, and QuIP\# \cite{tseng2024quip} achieve low-precision weights by minimizing layer-wise quantization error and optimizing parameters. Additionally, AWQ \cite{lin2023awq} and OWQ \cite{lee2023owq} consider the impact of activation outliers on weight quantization, resulting in improved performance for LLMs. OmniQuant~\cite{shao2023omniquant} further introduces learned weight clipping parameters to achieve lossless W4A16 quantization. While weight-only quantization can reduce the storage costs for modern inference systems, it has limited ability to decrease computational costs and provides minimal performance improvement for processing long sequences. Therefore, our work primarily focuses on achieving low-precision activation quantization.

\textbf{Weight-activation Quantization. }
Unlike weight-only quantization methods, weight-activation quantization techniques target both weights and activations, including the KV cache. As discussed in \cite{dettmers2022llm}, the biggest challenge in quantizing activations is handling outliers, which can be orders of magnitude larger than typical values. To address this, SmoothQuant \cite{xiao2023smoothquant} introduces an equivalent transformation strategy that partially shifts the quantization difficulty from activations to weights, achieving practical W8A8 solution. To further enhance the accuracy of quantized models, some works have explored fine-grained quantization strategies specifically tailored for activations. For example, KVQuant and WKVQuant \cite{yue2024wkvquant, hooper2024kvquant} focus on achieving fine-grained quantization for the KV cache by adopting per-token or per-channel quantization strategies. While these approaches can reduce memory usage, they are inefficient for computing. ZeroQuant~\cite{yao2022zeroquant} and Atom~\cite{zhao2024atom} apply per-channel quantization for activations, while RPTQ~\cite{yuan2023rptq} proposes a reordering strategy to non-uniformly group different parts of activations. Recent works~\cite{lin2024qserve, ashkboos2024quarot, liu2024spinquant} further employ channel-wise or even group-wise quantization for activations. However, none of these methods can efficiently execute on modern GPUs due to the misalignment between the granularity of quantization and the computational granularity of GPUs. In comparison, in this paper, we propose a fine-grained quantization strategy that aligns with the computational granularity of GPUs. Additionally, we utilize mixed-precision techniques to further improve the compression ratio of the LLM model, thereby enhancing both memory efficiency and computational performance.

\begin{figure}
    \centering
    \includegraphics[width=0.98\linewidth]{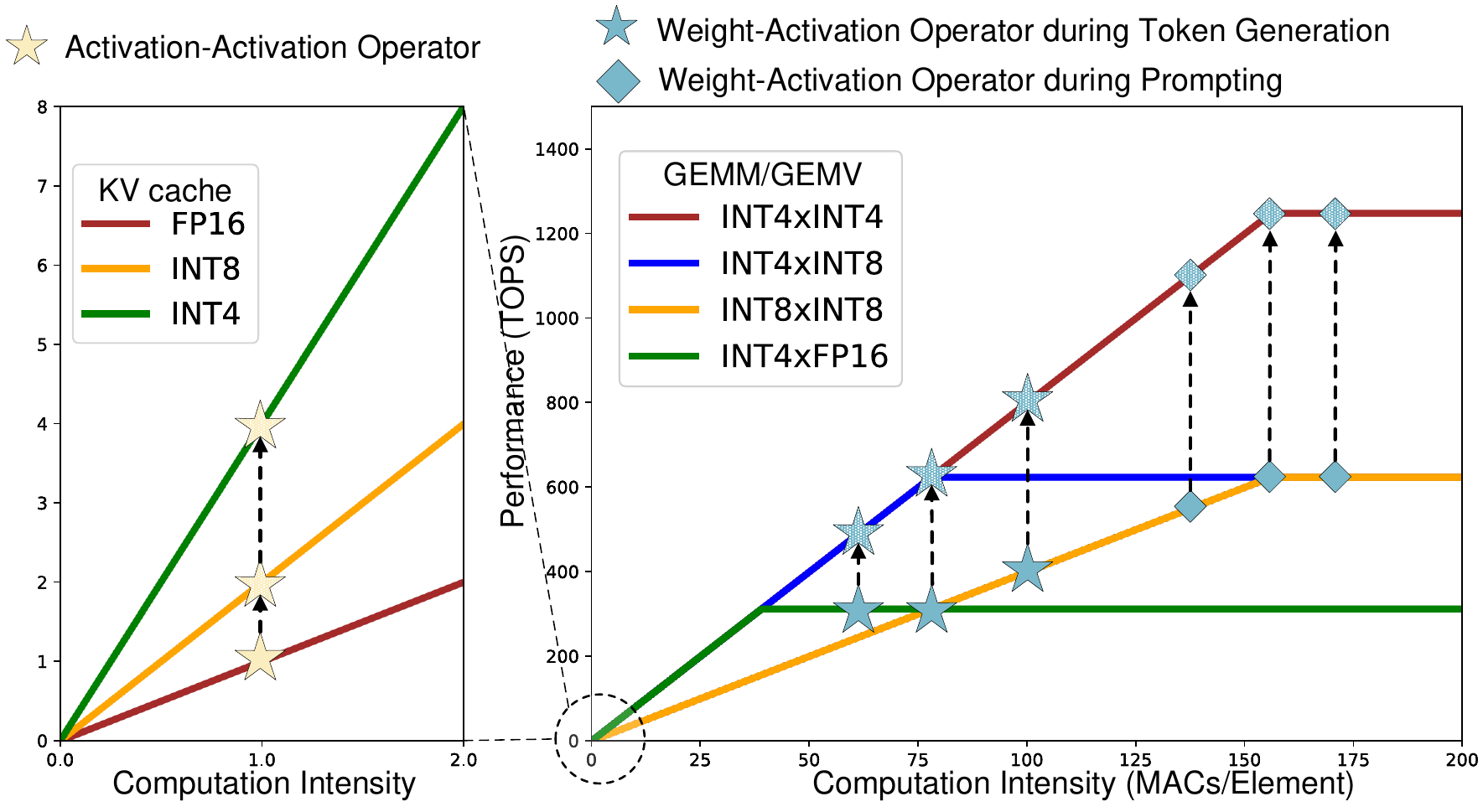}
    \caption{Roofline model analysis for activation-activation and weight-activation operators. The activation-activation operator is highly memory-bound, while the weight-activation operator becomes compute-bound with large-batch parallelism.}
    \label{fig:roofline}
\end{figure}

\subsection{Motivation}\label{sec:motivation-analysis}
Existing LLM serving systems~\cite{tensorrt-llm, rasley2020deepspeed} typically utilize modern GPUs, which are equipped with high-bandwidth memory (HBM) and powerful processing units such as high-throughput tensor cores. For instance, the A100 80G offers 80GB of HBM with 2.0TB/s bandwidth and can deliver up to 312 TFLOPS of FP16 tensor core performance for GEMM operations. In addition, modern GPUs support lower-precision computations, achieving 624 TOPS with INT8 tensor cores and 1248 TOPS with INT4 tensor cores. Modern GPUs also include CUDA cores for handling complex computations such as data conversion and permutation.

As depicted in Figure \ref{fig:roofline}, we employ the classic roofline model to evaluate the performance of LLM inference on GPUs. According to Figure \ref{fig:llm-execution}, we assess the performance of two types of operators, activation-activation operators and weight-activation operators, under different precisions. The computation intensity of the activation-activation operator is fixed at 1.0, making it memory-bandwidth bound. Therefore, implementing low-bit quantization for the KV cache can significantly enhance the throughput of activation-activation operators. Conversely, the computation intensity of weight-activation operators varies with batch size and inference phases. Thus, under large-batch parallelism, adopting low-precision activation quantization can further improve inference throughput. These observations motivate us to explore low-precision quantization for activations (including input activations and the KV cache) more thoroughly. Additionally, due to being bound by different characteristics, we will adopt different quantization strategies for input activation and KV cache,  as detailed in Section \ref{sec:activation-quant}.
\section{FMPQ: Fine-Grained Mixed-Precision Quantization}\label{sec:algorithm-design}


To address the challenges of activation quantization in post-training quantization, this section first analyzes the characteristics of activation distributions in LLMs and then proposes a mixed-precision quantization algorithm to achieve low-bit quantization of LLM activations. The proposed FMPQ effectively reduces the computational and storage costs of LLM inference, serving as the foundational enabler of our COMET inference framework.

\begin{figure}
    \centering
    \begin{subfigure}{0.48\linewidth}
        \centering
        \includegraphics[width=0.99\linewidth]{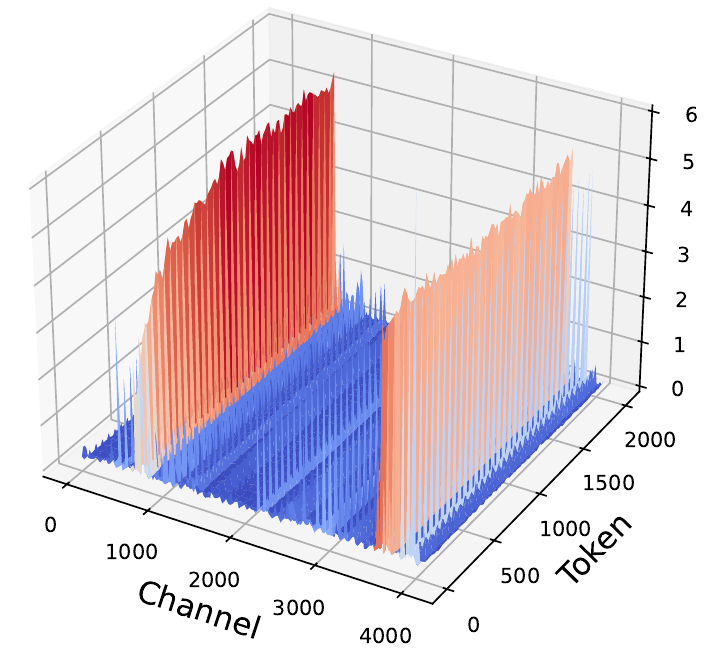}
        \caption{Layer10\_Attn\_K\_proj}
        \label{fig:k_outlier}
    \end{subfigure}
    \centering
    \begin{subfigure}{0.48\linewidth}
        \centering
        \includegraphics[width=0.99\linewidth]{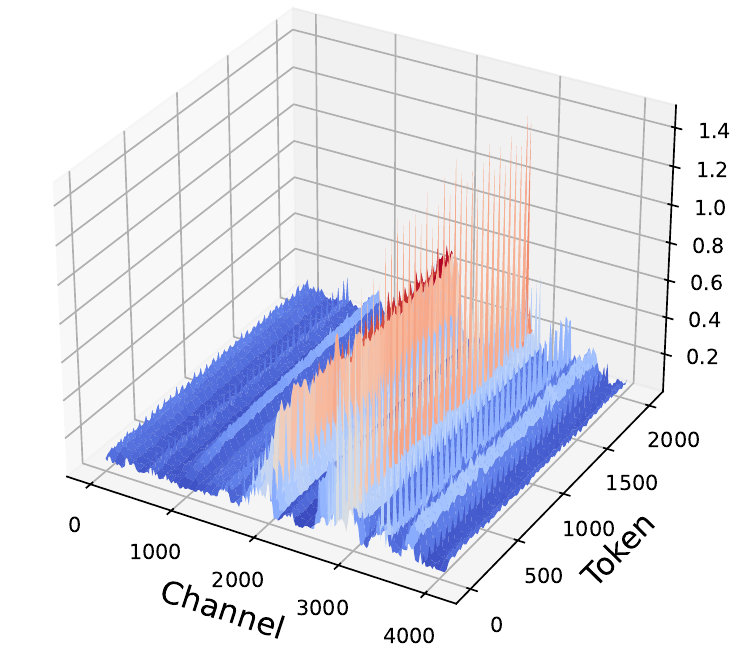}
        \caption{Layer29\_Attn\_O\_proj}
        \label{fig:o_outlier}
    \end{subfigure}
    \caption{Activation distributions in LLaMA-7B. Outliers are present in certain channels of activations, leading to the ineffectiveness of conventional quantization strategies.}
    \label{fig:outlier-distribution}
\end{figure}

\subsection{Analysis of Activation Distribution}\label{sec:algorithm-motivation}

Unlike small-scale neural networks~\cite{vaswani2017attention, dosovitskiy2020image, liu2021swin}, a distinctive characteristic of LLMs is the pervasive presence of outliers. Once a model exceeds a certain scale (in practice, around 6 billion parameters), a small cluster of hidden state features (usually less than $1\%$) exhibits magnitudes significantly larger than the rest~\cite{dettmers2022llm}. These emergent large-magnitude features are referred to as outliers, and they have magnitudes that can exceed typical hidden state values by tenfold or even a hundredfold. When using traditional min-max quantization strategies on activation tensors, these outliers can cause typical hidden state values to be quantized to zero, resulting in significant accuracy degradation. These outliers are so crucial to the representational capacity of LLMs that discarding them is not feasible~\cite{guo2023olive}.

Fortunately, we have conducted extensive experimental analysis on current typical LLMs, such as LLaMA, and found that these outliers typically distribute within specific feature dimensions. As illustrated in Figure \ref{fig:outlier-distribution}, the outliers in LLaMA-7B only exist in certain activation channels. Therefore, an intuitive approach is to separately quantize outliers and normal values, quantizing the outliers with high precision and the normal values with low precision.

\subsection{The FMPQ Algorithm}\label{sec:activation-quant}

\begin{figure}
    \centering
    \includegraphics[width=0.98\linewidth]{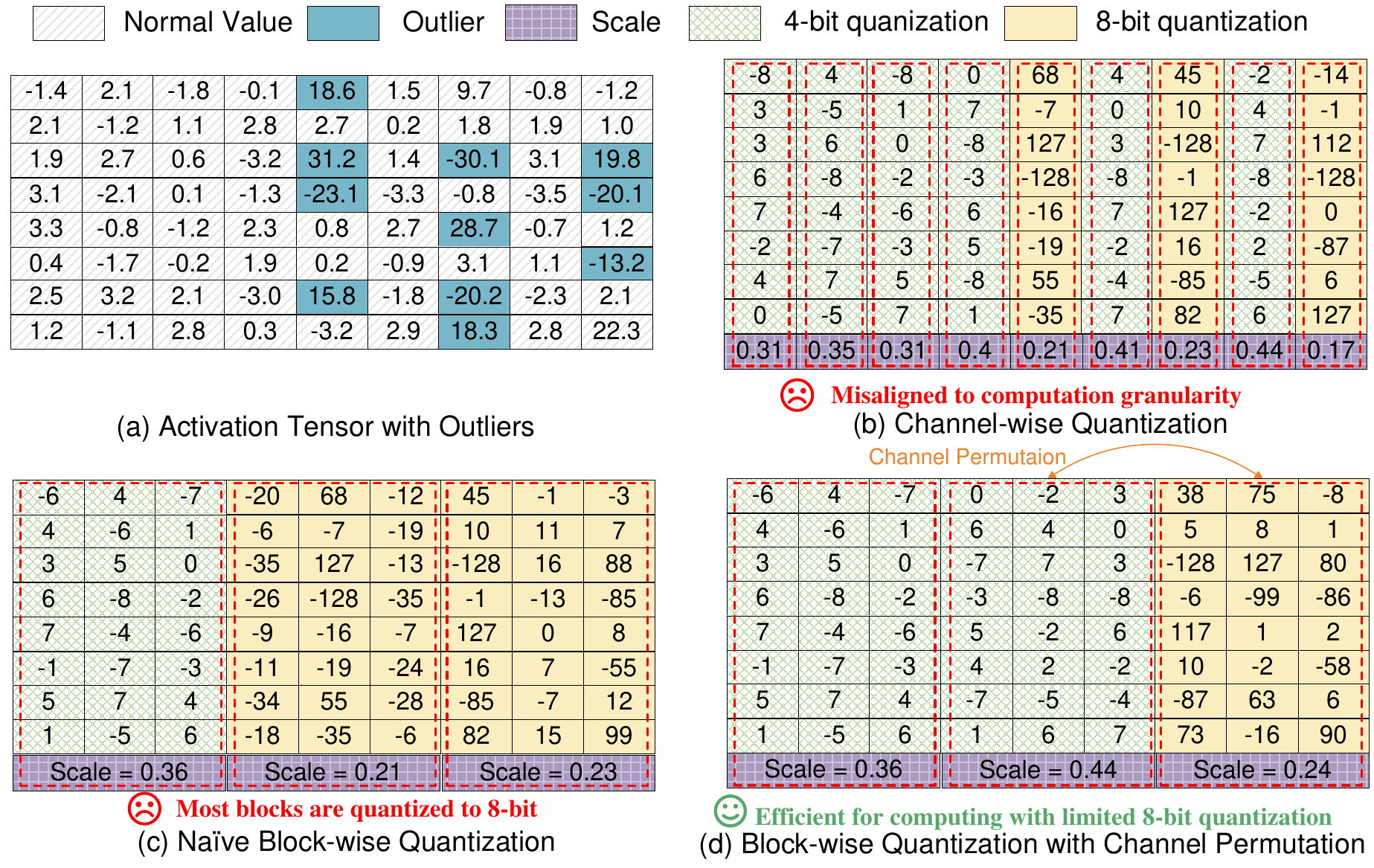}
    \caption{The algorithm design of FMPQ.}
    \label{fig:algorithm-design}
\end{figure}

With the above observation, we can apply a fine-grained mixed-precision quantization algorithm that partitions the normal values and outliers into different parts, and quantizes them with different precision. For example, we quantize normal values to 4-bit, while outliers to 8-bit. However, as demonstrated in Figure \ref{fig:algorithm-design}b, previous channel-wise quantization algorithms~\cite{yao2022zeroquant, zhao2024atom} cannot ensure the efficiency of the quantized model due to their misalignment with hardware computation granularity.


In fact, the minimum computation granularity on modern GPUs is fixed. For instance, the FP16 tensor core on A100 has a minimum computation granularity of $64 \times 64 \times 32$ ($m \times n \times k$). This hardware constraint necessitates that the granularity of quantization is an integer multiple of the minimum computation granularity, to maximize the utilization of tensor core. Consequently, prior channel-wise or irregular cluster quantization strategies~\cite{yao2022zeroquant, yuan2023rptq}, which fail to align well with hardware computation granularity, are inefficient. On the other hand, overly fine-grained quantization strategies (e.g., tile-wise quantization) introduce significant software scheduling overhead. Therefore, as illustrated in Figure~\ref{fig:algorithm-design}c, we adopt a block-wise mixed-precision quantization method, which partitions the activation tensor only along the channel dimension. Specifically, we divide the activation tensor into multiple sub-tensors with a channel size of $k$, referring to these sub-tensors as blocks. In practice, we find that setting $k$ to 128 ensures sufficient tensor core utilization without excessive accuracy loss in the quantized LLM.

Since outliers are distributed across multiple channels, any block that contains outliers will require 8-bit quantization. This results in a large number of blocks being quantized to 8-bit, which limits the benefits, as depicted in Figure~\ref{fig:algorithm-design}c. To address this issue, we employ a channel permutation strategy, which is commonly used in LLM pruning~\cite{zhang2024plug, pool2021channel}. As illustrated in Figure \ref{fig:algorithm-design}d, we first identify channels with outliers through data sampling and then use the permutation strategy to cluster these channels into a single block. To maintain computational equivalence, the corresponding positions in the weight matrix also need to be permuted. In this way, only a small portion of blocks (less than $20\%$) need 8-bit quantization, while the majority of activations can be quantized to 4-bit, enabling W4A4 GEMM. In fact, prior works~\cite{li2023sparse, yuan2023rptq} have already optimized the channel permutation operator, and according to our evaluation, channel permutation accounts for only $0.7\%$ of the overall runtime.


While the proposed channel permutation enhanced block-wise quantization works well for input activations, the KV cache requires a different quantization strategy. As analyzed in Figure~ \ref{fig:roofline}, the activation-activation operator which KV cache works for is highly memory-bound. Therefore, KV cache is better suited for low-bit quantization without considering the quantized granularity. Furthermore, as the KV cache is more amenable to quantization than input activations~\cite{zhao2024atom, hooper2024kvquant}, we can apply a full 4-bit quantization to the KV cache. Specifically, we use a channel-wise 4-bit quantization strategy for the KV cache. The experimental results in Section \ref{sec:alg-evaluation} demonstrate that this approach has a negligible impact on accuracy. By combining the quantization for input activation and KV cache, the FMPQ algorithm provides a practical path for high-performance LLM serving. 

\begin{figure*}
    \centering
        \begin{subfigure}{0.24\linewidth}
        \centering
        \includegraphics[width=0.99\linewidth]{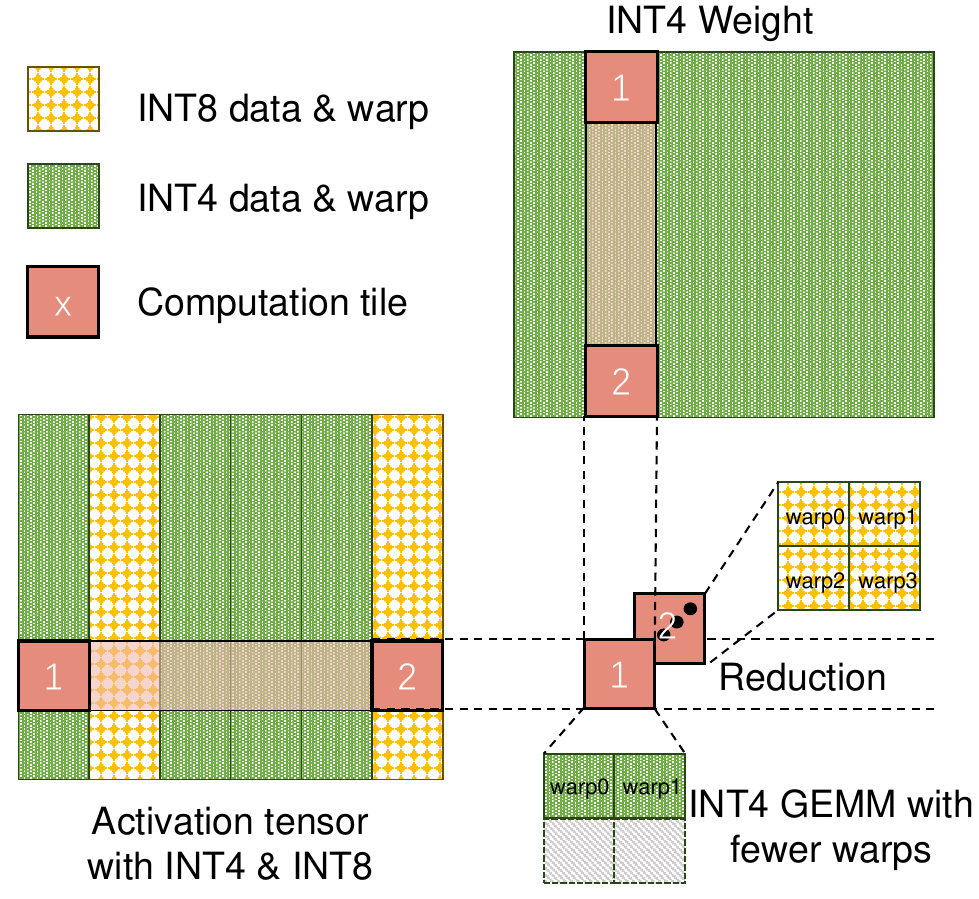}
        \caption{Tile-based computing with mixed-precision.}
        \label{fig:compute-process}
    \end{subfigure}
    \centering
    \begin{subfigure}{0.30\linewidth}
        \centering        \includegraphics[width=0.99\linewidth]{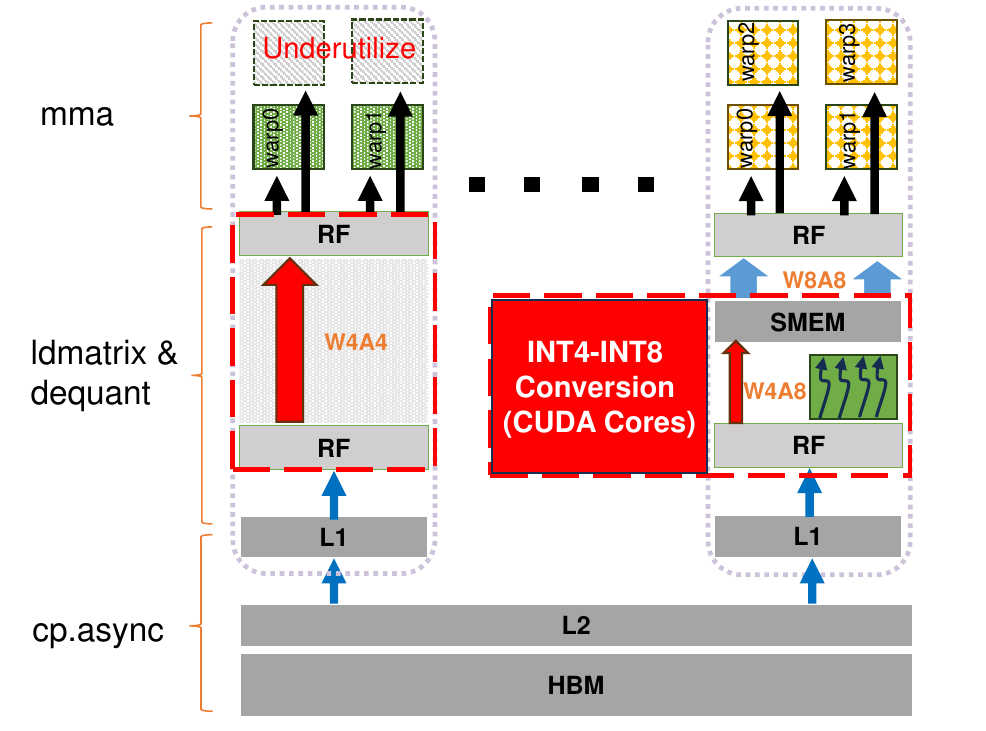}
        \caption{Instruction issued for different SMs with different precision.}
        \label{fig:kernel-overview}
    \end{subfigure}
    \centering
    \begin{subfigure}{0.40\linewidth}
        \centering
        \includegraphics[width=0.99\linewidth]{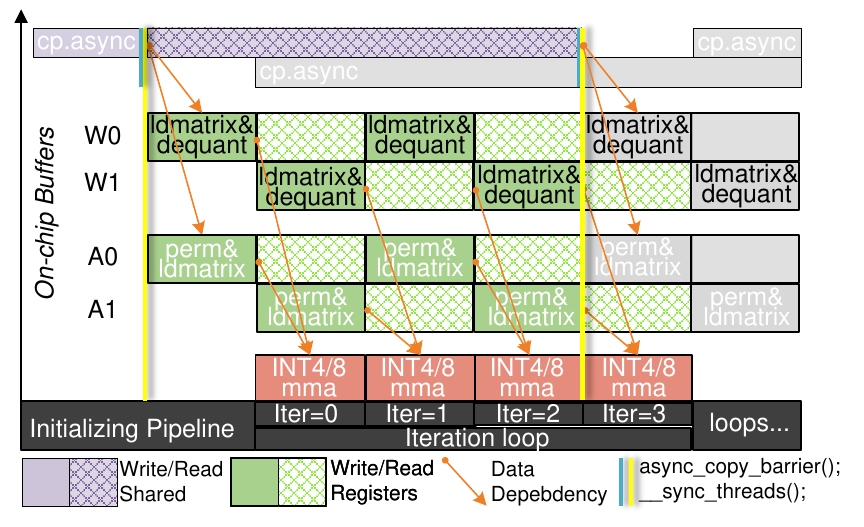}
        \caption{SIMT-enhanced software pipeline. }
        \label{fig:execution-pipeline}
    \end{subfigure}
    \caption{The design overview of the COMET-W4Ax kernel. (a) illustrates the tile-based GEMM computing with mixed-precision encoding. (b) presents the computing procedure when issuing W4A4 and W4A8 GEMM instructions simultaneously. (c) shows that we can effectively hide the overhead by software pipeline.}
    \label{fig:kernel-design}
\end{figure*}

\section{COMET-W4Ax: Kernel Design}\label{sec:COMET-kernel}


The proposed FMPQ algorithm can significantly reduce the storage and computing costs in LLM inference. However, existing LLM serving systems~\cite{tensorrt-llm, llama.cpp} lack support for direct mixed-precision tensor load-and-store and W4Ax computing. Thus, in this section, we design a highly optimized W4Ax kernel for COMET by tackling two main challenges: (1) the additional overhead of data management with mixed-precision encoding, and (2) load imbalance induced by varied W4A4 and W4A8 GEMM operations.

\subsection{Design Overview}\label{sec:design-overview}
GEMM computations on GPUs are performed at the tile granularity. Figure \ref{fig:compute-process} illustrates the tile-based GEMM computing in COMET using mixed-precision values. With FMPQ as the algorithmic enabler, the activation tensor is divided into multiple blocks with different precision settings. For example, the green block is quantized to 4-bit, while the yellow block is 8-bit. A block usually contains multiple tiles, and each tile invokes one thread block (TB) to compute. Additionally, we utilize the reduction operator to accumulate the compute results of different tiles across multiple TBs.

Figure \ref{fig:kernel-overview} shows the behavior of the COMET-W4Ax kernel when computing different precision tiles among different SMs at the same time. During the kernel execution, data must first be loaded from global memory into shared memory and then dispatched to each SM's tensor core for computation. As presented in Section \ref{sec:software-pipeline}, we use a software pipeline to concurrently handle data loading and GEMM computation. COMET-W4Ax primarily involves two types of GEMM computations during kernel execution: W4A4 and W4A8. The W4A4 computation can directly leverage the \texttt{mma} instructions, whereas the W4A8 GEMM requires additional data conversion. Specifically, we utilize CUDA cores to efficiently convert INT4 to INT8 data formats and store the converted results directly in shared memory to support W4A8 GEMM. We present the detailed optimization on data conversion in Section \ref{sec:mixed-management}. Additionally, we notice that when issuing GEMM computation instructions for W4A4 and W4A8 to different SMs simultaneously, the computational resources of the tensor cores used for W4A4 are not fully utilized. Despite the INT4 tensor cores offering $2 \times$ higher throughput than INT8 tensor cores, the SMs executing W4A4 computations have to wait for other SMs to achieve synchronization, leading to significant idle times and low resource utilization. In Section \ref{sec:sm-scheduling}, we address this issue using fine-grained SM scheduling. 


\subsection{SIMT-enhanced Software Pipeline}\label{sec:software-pipeline}

Besides traditional GEMM computations, there are extensive data reformatting and repositioning operations (e.g., dequantization and permutation) involved. These operations introduce additional overhead. Therefore, we integrate the overhead of these operations into the traditional software pipeline, effectively hiding the costs of dequantization and permutation. Multiple stages are required to load mixed-precision data from global memory into the tensor core for GEMM computation. Specifically, the process begins by using \texttt{cp.async} to load data from the global memory into shared memory or distributed registers. Additional instructions, such as permutation and dequantization, are then employed to adjust the data location and format. Finally, the data is mapped to the corresponding tensor cores where \texttt{mma} instructions are invoked for computation. Note that, different tiles use INT4 or INT8 instructions for \texttt{mma} computation of W4A4 and W4A8, respectively.

Specifically, COMET utilizes a two-level overlapping strategy to pipeline the memory access and computation stages, achieving efficient GEMM computation (presented in Figure \ref{fig:execution-pipeline}). On the one hand, it hides off-chip memory loads within the data transformation and tensor core computation phases. On the other hand, it leverages double buffering within the GPU to conceal the overhead of tensor core computation and data transfer/transformation. Specifically, COMET uses two shared memory buffers to store different data tiles (storing $A_0$ and $W_0$ in $buffer_0$, and $A_1$ and $W_1$ in $buffer_1$). In iteration 0, the tensor core loads $A_0$ and $W_0$ from $buffer_0$ and performs the corresponding GEMM computation. Depending on the data format loaded, the tensor core invokes different \texttt{mma} instructions (INT4 or INT8). Simultaneously, $buffer_1$ loads the relevant data for $A_1$ and $W_1$ from global memory and decides whether permutation or dequantization adjustments are necessary. In iteration 1, the tensor core loads data from $buffer_1$ while $buffer_0$ prefetches data from global memory. In this manner, the data loading and the tensor core computation effectively overlap.

Given the complexity of the pipeline, it is necessary to introduce a set of thread synchronizations and memory access barriers to ensure the correctness of model execution. COMET inserts a series of synchronization and barrier instructions to maintain correctness. Specifically, the \texttt{async\_copy\_barrier} effectively ensures that all necessary data is ready in shared memory before the execution of the next iteration loop, while \texttt{sync\_threads} ensures that all threads are synchronized before the next iteration loop begins. Moreover, to prevent all mixed-precision tiles from being written back before completing GEMM computation and reduction, COMET explicitly inserts a thread-block level synchronization to ensure that all threads have completed their work and that all tile reductions are performed. 


\begin{figure}
    \centering
    \includegraphics[width=0.98\linewidth]{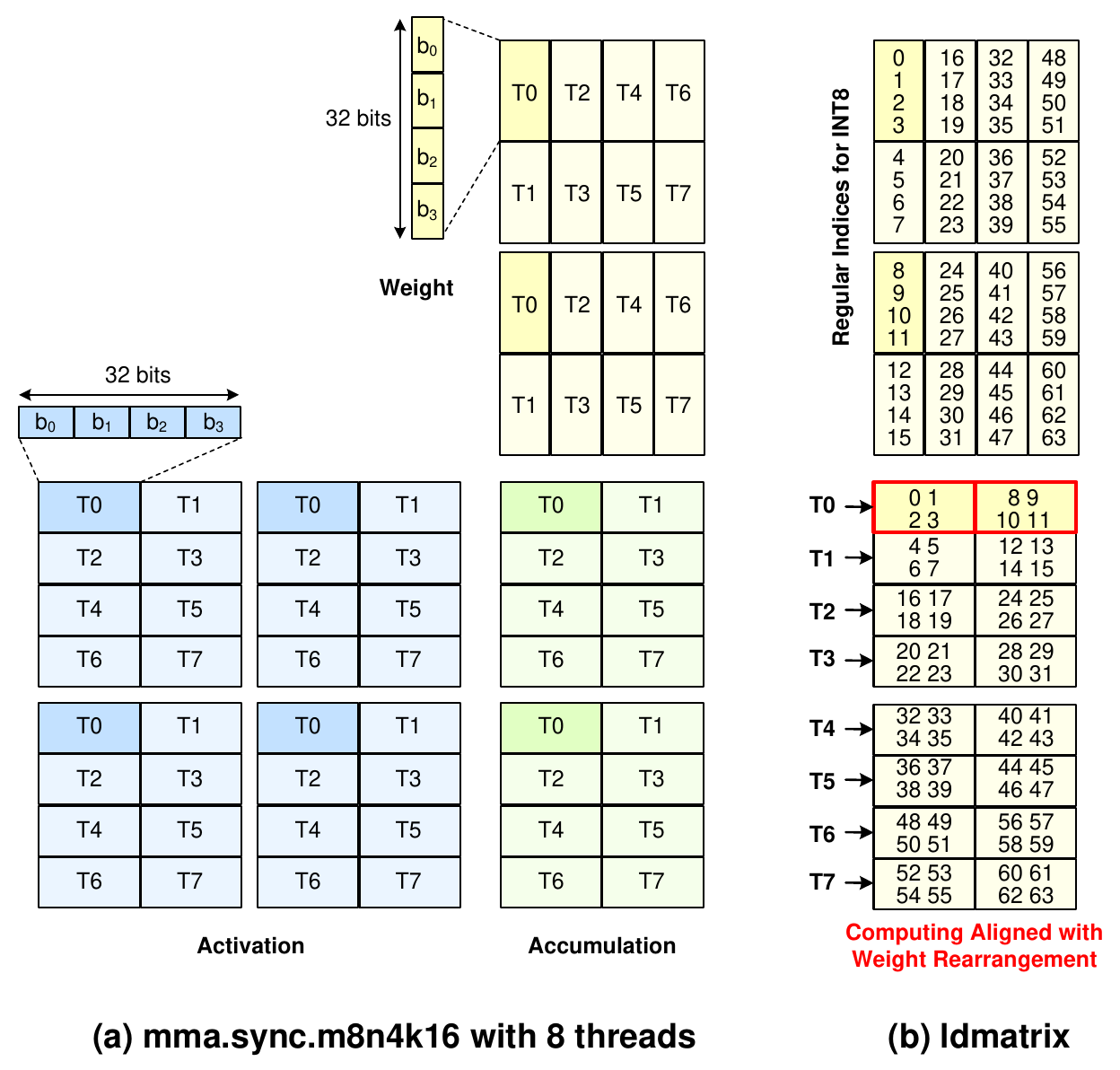}
    \caption{Weight interleave for fast access of W4A8 GEMM. }
    \label{fig:weight-rearrangement}
\end{figure}

\subsection{Mixed-precision Data Management}\label{sec:mixed-management}
The COMET-W4Ax kernel involves GEMM computations of different precisions, including W4A4 and W4A8 GEMM computations. However, the GPU tensor core only supports GEMM computations with data in the same format. Therefore, to optimize the efficiency of W4A8 computations, we need to implement efficient data layout and data conversion strategies.

\textbf{Weight Interleave for W4A8 GEMM. }
Before performing actual computations on the GPU, all operands must be loaded from shared memory into the corresponding registers for each tensor core (\texttt{ldmatrix}). Subsequently, a thread scheduler is used to load the required data into the tensor core for the corresponding GEMM computation (\texttt{mma}). For W4A8 GEMM, the tensor core performs computations using INT8 tensor cores to execute W8A8 computations. The tensor core relies on the \texttt{mma.m16n8k32} instruction to perform INT8 GEMM computations, involving 32 threads. Given the complexity of the \texttt{mma} execution, Figure \ref{fig:weight-rearrangement}a shows a simplified example of \texttt{mma.m8n4k16} GEMM computation which only uses 8 threads. As it illustrates, each thread needs to load activation data four times and weight data two times, then conduct the multiply-accumulate (MAC) operations two times for a na\"ive W8A8 GEMM. 

In existing GEMM optimization frameworks, as long as the format of the loaded data matches the computation data format, programmers can directly use the \texttt{ldmatrix} instruction to ensure the loaded data corresponds to the required computation data. However, directly transferring 4-bit weights in 8-bit to the tensor core can cause shared memory conflicts. As shown in Figure \ref{fig:weight-rearrangement}b, the weights that a single thread (e.g., thread 0) needs to load are non-contiguous. To avoid shared memory conflicts, we need to rearrange the layout of the weight data. Specifically, thread 0 uses addresses 0-3 and 8-11 instead of contiguous 0-7 values to align the computation and memory access granularity. This weight interleave strategy not only avoids shared memory conflicts but also reduces the usage of \texttt{ldmatrix} instructions. We achieve the necessary data load in one instruction, compared to the two load instructions required for the INT8 scenario.

\begin{figure}
    \centering
    \includegraphics[width=0.98\linewidth]{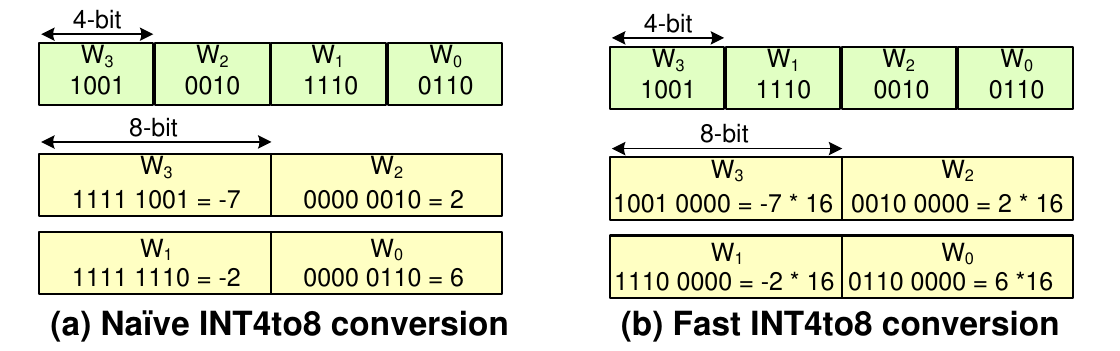}
    \caption{Fast data conversion from INT4 to INT8.}
    \label{fig:data-conversion}
\end{figure}

\textbf{Fast INT4to8 Conversion. }
Since tensor cores only support GEMM computations with data in the same format, it is necessary to convert INT4 data to INT8 before performing GEMM computation. Data format conversion can only be executed on CUDA cores. However, in modern GPUs, there is a significant performance gap between CUDA cores and tensor cores. For example, on the A100, the computational throughput of the INT8 tensor core is $32 \times$ higher than that of the CUDA core. Therefore, we must implement a fast INT4 to INT8 data format conversion, making it a stepping stone for performance enhancement rather than a bottleneck.

Suppose we pack four 4-bit weight values into a single 16-bit data unit and store them sequentially. Figure \ref{fig:data-conversion}a illustrates a na\"ive strategy for converting INT4 to INT8. In this process, data position adjustments and sign extensions are required. For example, in the conversion of $W_1$ and $W_0$, we first need to shift the data position of $W_1$ from bits 4-7 to 8-11, and then perform sign extension on both $W_0$ and $W_1$. Unfortunately, the PTX ISA~\cite{PTXISA} does not support 4-bit shift operations or the corresponding sign extension operations, requiring us to rely on multiple instructions to complete the data conversion. Overall, each conversion can take up to 10 instructions to complete the necessary processing. 

To mitigate the overhead of data position adjustments and sign extension, we propose two novel strategies for fast value conversion. First, we use a location switch to swap the storage positions of certain data. As shown in Figure \ref{fig:data-conversion}b, we swap the positions of $W_1$ and $W_2$, enabling a rapid conversion to the corresponding 8-bit data format addresses. Additionally, we employ zero extension instead of sign extension for data conversion. Unlike INT4 sign extension which relies on multiple instructions to achieve, zero extension, on the other hand, can be achieved directly with a single zero-padding instruction. During zero extension, each value is effectively multiplied by 16, so we only need to divide by 16 in the scaling parameter to ensure consistency in the computation results. In comparison to na\"ive conversion, the overhead with our proposed strategy for each value is reduced to just 2 instructions for each conversion. This significantly improves the efficiency of data conversion.

\subsection{Fine-Grained SM Scheduling}\label{sec:sm-scheduling}

\begin{figure*}
    \centering
    \includegraphics[width=0.8\linewidth]{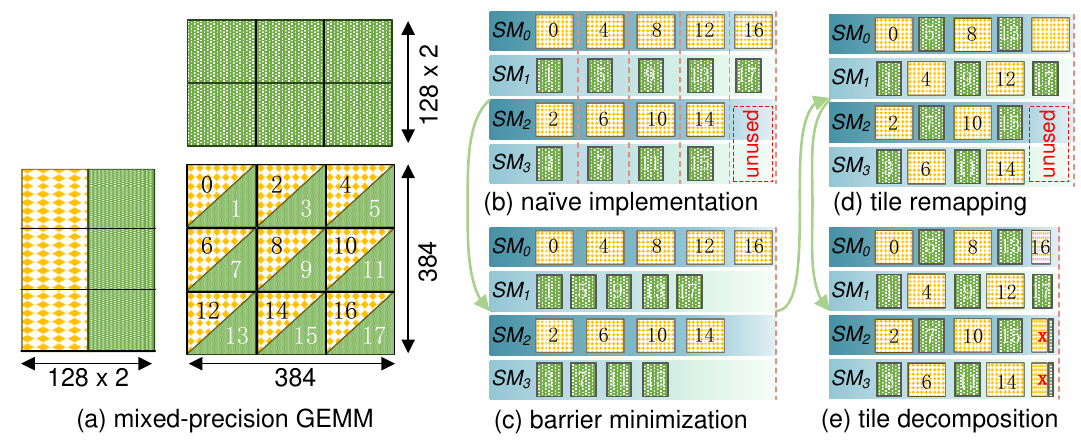}
    \caption{The proposed fine-grained SM scheduling. Instead of using a na\"ive implementation, we further adopt tile remapping and tile decomposition to greatly improve the tensor core utilization.}
    \label{fig:sm-scheduling}
\end{figure*}

Given our implementation of fine-grained mixed-precision quantization, different blocks in the COMET system may require different precision \texttt{mma} instructions for computation. Figure \ref{fig:sm-scheduling}a illustrates such a scenario on a hypothetical GPU with four streaming multiprocessor cores. Specifically, we set the block size $k=128$. Consider a GEMM operation of size $256 \times 256 \times 384$, which includes two activation blocks in INT4 and INT8 formats, we divide the computation into 18 tiles. The shape of each tile is $128 \times 128 \times 128$. During the GEMM computation process, every two consecutive tiles compute the W4A8 and W4A4 GEMM kernels respectively.

W4A4 GEMM computations are stalled by inserted barriers to wait for the W4A8 GEMM and ensure the correctness of the computation, even with $2 \times$ higher throughput. As illustrated in Figure \ref{fig:sm-scheduling}b, during the computation process, the INT4 \texttt{mma} instruction needs to wait for the INT8 \texttt{mma} instruction to complete, significantly reducing SM utilization. For example, $SM_1$ and $SM_3$ must wait for $SM_0$ and $SM_2$ to complete their INT8 \texttt{mma} computations in each iteration to achieve synchronization. Additionally, as the red dotted box shows, the mismatch between the number of tiles and the number of SMs leads to some SMs being idle in the final iteration, further lowering the overall execution efficiency. To address this, we propose three strategies for fine-grained SM scheduling, achieving load balance across different SMs.

\textbf{Synchronization Barrier Minimization. }
It is not necessary to insert synchronization barriers for every \texttt{mma} iteration.
We notice that synchronization is only necessary after all iterations are completed, as illustrated in Figure \ref{fig:sm-scheduling}c. Frequent synchronization barriers introduce additional communication overhead between SMs. Therefore, to minimize interaction between SMs, we should reduce the insertion of synchronization barriers as much as possible while ensuring computational correctness. The only synchronization required between SMs is the one that before writing the accumulation data back to memory. However, the performance gains from minimizing synchronization barriers are limited. Specifically, as one can notice, the SMs executing INT4 \texttt{mma} ($SM_1$ and $SM_3$) always wait for $SM_0$ and $SM_2$ to complete their corresponding computations before synchronization. As a result, the utilization of $SM_1$ and $SM_3$ remains low compared to $SM_0$ and $SM_2$. 

\textbf{Tile Remapping. }
Mapping all INT4 and INT8 \texttt{mma} computations to fixed SMs is unnecessary. To address this, we propose an effective tile remapping strategy that adjusts the mapping relationship between tiles and SMs to achieve effective load balancing across different SMs. As illustrated in Figure \ref{fig:sm-scheduling}d, we distribute the INT4 and INT8 \texttt{mma} computations as evenly as possible across all SMs. This ensures that each SM has a balanced computational load, reducing the overall kernel execution time. However, due to the mismatch between the number of divided tiles and the number of SMs, some SM resource wastage still occurs. Fine-grained tuning is required to further enhance the execution efficiency of mixed-precision GEMM.

\textbf{Tile Decomposition. }
To achieve better alignment between tile division and SM resources, the most straightforward approach is to implement finer-grained tile division. However, this finer-grained blocking factor is less cache and scratchpad efficient, potentially negating any practical performance improvements. In this paper, we reexamine the binding relationship between tiles and SMs. The tile-to-SM binding is traditionally one-to-one, meaning that at any given time, a tile is mapped to a single SM. However, through effective scheduling, we can achieve a one-to-many binding~\cite{osama2023stream} between tiles and SMs. Specifically, we implement a task-stealing mechanism, where idle SMs can steal computational tasks from nearby busy SMs to balance the workload, as illustrated in Figure \ref{fig:sm-scheduling}e. For instance, when $SM_0$ and $SM_1$ are processing tile-16 and tile-17 respectively, $SM_2$ and $SM_3$ remain idle. During this period,  $SM_2$ and $SM_3$ load data from shared memory and assist $SM_0$ and $SM_1$ with portions of their computations. This mechanism enhances the utilization of SMs and reduces the overall execution time. 
\section{COMET: System Implementation}

We have implemented a practical mixed-precision serving framework, COMET, for high-performance LLM inference. COMET provides an easy-to-use Python interface for programmers. COMET leverages KV cache management optimizations from vLLM~\cite{kwon2023efficient}. Additionally, we introduce the W4Ax kernel to support mixed-precision matrix multiplication $O = WX$, where
$W$ is the weight matrix and $A$ is the activation matrix. The COMET-W4Ax kernel has been implemented by incorporating an additional 7,000 lines of C++ and CUDA code based on TensorRT-LLM~\cite{tensorrt-llm} and CUTLASS~\cite{cutlass}. The W4Ax kernel can be compiled as a standalone \texttt{.so} dynamic library. We also provide a set of C++ APIs so that programmers can easily integrate the optimized kernel in existing inference systems such as TensorRT-LLM and llama.cpp~\cite{llama.cpp}. To further simplify usage, we use \texttt{pybind} to bind these interfaces to Python. This allows programmers to quickly and easily integrate our kernel into Python-based frameworks, such as Pytorch and Huggingface~\cite{jain2022hugging}.



In our implementation, the tile size presented in Figure \ref{fig:compute-process} is set as $128 \times 128 \times 128$  ($ m\times n\times k $) for most cases. Furthermore, we set the warp shape as $64 \times 64 \times 128$ for INT4 tensor core, while $64 \times 64 \times 64 $ for INT8 tensor core. As a result, the number of warps issued by one W4A4 tile is only half of the W4A8 tile. These configurations are selected based on extensive preliminary evaluation and are intended to optimize the computational throughput while balancing memory access patterns and processing power. By maintaining these fixed configurations across all our experiments, we aim to clearly demonstrate the performance improvement from detailed kernel optimization. The fixed configuration allows us to isolate the effects of other variables and focus on analyzing the impact of different optimization strategies and algorithmic adjustments. The chosen shapes reflect the typical dimensions used in high-performance computing scenarios and are well-suited to the architecture of modern GPUs, facilitating efficient parallel processing and minimizing latency.
\section{Evaluation}\label{sec:evaluation}

\begin{table*}
\caption{WikiText perplexity of quantized LLaMA, LLaMA-2 and LLaMA-3 models. The lower is the better.  
}
\label{tab:ppl_results}
\centering
\small
\setlength{\tabcolsep}{6.5pt}
\renewcommand{\arraystretch}{1.0}
\resizebox{0.8\textwidth}{!}{
\begin{tabular}{l l c c c c c c c c c }
\toprule 
        &  &  \multicolumn{4}{c}{LLaMA} & \multicolumn{3}{c}{LLaMA-2} & \multicolumn{2}{c}{LLaMA-3}\\
    \cmidrule(lr){3-6}\cmidrule(l){7-9}\cmidrule(l){10-11}
   Precision & Method & 7B & 13B & 30B & 65B & 7B & 13B & 70B & 8B & 70B \\
    \hline
  FP16  & - & 5.68 & 5.09 & 4.10 & 3.56 & 5.12 & 4.57 & 3.12 & 6.14 & 2.86 \\
    \hline
  W8A8 & SmoothQuant & 5.78 & 5.19 & 4.23 & 5.37 & 5.54 & 4.95 & 3.36 & 6.28 & 2.99\\
    \hline 
  \multirow{3}*{W4A16} & GPTQ & 6.13 & 5.40 & 4.48 & 3.83 & 5.83 & 5.13 & 3.58 & 7.02 & 3.44 \\ 
   & AWQ & 6.08 & 5.34 & 4.39 & 3.76 & 6.15 & 5.12 & 3.54 & 7.09 & 3.40\\
   & Omniquant & 5.86 & 5.21 & 4.25 & 3.71 & 5.74 & 5.02 & 3.47 & 6.81 & 3.29 \\
  \gr W4Ax & \textbf{FMPQ} & 5.88 & 5.29 & 4.27 & 3.78 & 5.71 & 5.10 & 3.48 & 6.88 & 3.36 \\ 
  W4A4 & Omniquant & 11.26 & 10.87 & 10.33 & 9.17 & 14.26 & 12.30 & 9.93 & 14.27 & 9.75 \\
    \hline
    W4A8 KV4 & QoQ & 5.93 & 5.28 & 4.34 & 3.83 & 5.75 & 5.12 & 3.52 & 6.89 & 3.38 \\
  \gr \textbf{W4AxKV4} & \textbf{FMPQ} & 5.95 & 5.32 & \textbf{4.31} & \textbf{3.82} & \textbf{5.73} & 5.19 & 3.56 & 6.91 & 3.41\\ 
    \bottomrule
\end{tabular}
}
\end{table*}

\begin{table*}
\caption{Zero-shot accuracy evaluation on five common sense tasks for LLaMA-3 family models.
}
\label{tab:zero_shot_results}
\centering
\small
\setlength{\tabcolsep}{6.5pt}
\renewcommand{\arraystretch}{1.0}
\resizebox{0.8\textwidth}{!}{
\begin{tabular}{c l l c c c c c c}
\toprule 
        &  & & \multicolumn{6}{c}{Zero-shot Accuracy $\uparrow$} \\
    \cmidrule(lr){4-9}
   Size & \#Configuration & Method & PIQA & ARC-e & ARC-c & HellaSwag & Winogrande & Avg. \\
    \hline
    \multirow{4}*{8B} & FP16 & Full Precision & 79.9 & 80.1 & 50.4 & 60.2 & 72.8 & 68.6 \\
    & W8A8 & SmoothQuant & 79.5 & 79.7 & 49.0 & 60.0 & 73.2 & 68.3 \\
    & W4A16 & Omniquant & 78.4 & 77.9 & 48.5 & 58.8 & 72.7 & 67.2 \\
    & W4A8 KV4 & QoQ & 77.1 & 77.2 & 47.8 & 57.6 & 72.0 & 66.3 \\
   \gr &  \textbf{W4AxKV4} & \textbf{FMPQ} & {77.5} & 76.7 & 47.5 & \textbf{58.9} & {72.1} & {66.5} \\
    \hline
    \multirow{4}*{70B} & FP16 & Full Precision & 82.4 & 86.9 & 60.3 & 66.4 & 80.6 & 75.3 \\
    & W8A8 & SmoothQuant & 82.2 & 86.9 & 60.2 & 66.3 & 80.7 & 75.3 \\
    & W4A16 & Omniquant & 82.7 & 86.3 & 59.0 & 65.7 & 80.9 & 74.9 \\
    & W4A8 KV4 & QoQ & 81.4 & 85.7 & 58.4 & 64.9 & 79.9 & 74.0 \\
   \gr &  \textbf{W4AxKV4} & \textbf{FMPQ} & {82.5} & 85.2 & 58.3 & {65.0} & 79.6 & {74.1} \\
    \bottomrule
\end{tabular}
}
\end{table*}

\subsection{Experimental Setup}\label{sec:evaluation-setup}

\textbf{Algorithm. }
The proposed quantization algorithm, FMPQ, is implemented using HuggingFace on top of PyTorch. We employ block-wise mixed-precision (INT4 and INT8) quantization for activations and channel-wise asymmetric INT4 group quantization for the KV cache. Additionally, we adopt the algorithm in ~\cite{shao2023omniquant} to achieve 4-bit weight quantization. We use the term "\textbf{W4AxKV4}" to denote the configurations we adopt. Note that most of the computations are conducted on W4A4KV4 cases. 


\textbf{System. }
We evaluate the performance of the COMET inference system at two different levels: kernel-level benchmarking and end-to-end inference performance. All performance evaluations are conducted on the NVIDIA A100-80GB-SXM4 platform with CUDA 12.1. Our primary focus is on the performance of linear layers within LLMs during the token generation phase. GPU kernel performance is measured using NVIDIA Nsight Compute~\cite{nsight}. End-to-end inference throughput is measured using NVIDIA Nsight Systems~\cite{nsight-system}. For baseline systems, we use TensorRT-LLM (TRT-LLM) v0.10.0 to perform inference evaluations under different configurations.




\subsection{Algorithm Evaluation}\label{sec:alg-evaluation}

\textbf{Algorithm Benchmarks. }
We compare our proposed FMPQ algorithm with other baselines on the LLaMA-1~\cite{touvron2023llama}, LLaMA-2~\cite{touvron2023llama2} and LLaMA-3 family models. Following previous literature settings~\cite{frantar2022gptq, ashkboos2024quarot, chee2024quip, shao2023omniquant, liu2024spinquant}, we evaluated FMPQ-quantized models on language modeling and zero-shot tasks. Specifically, we evaluated the perplexity of quantized models on WikiText2~\cite{merity2016pointer}, and evaluated the zero-shot accuracy on PIQA~\cite{bisk2020piqa}, ARC~\cite{clark2018think} (including ARC-e and ARC-c), HellaSwag~\cite{zellers2019hellaswag} and WinoGrande~\cite{sakaguchi2021winogrande} with lm\_eval~\cite{gao2021framework}.

\textbf{Algorithm Baselines. } We compared FMPQ with widely used PTQ LLM quantization algorithms, including weight-only and weight-activation quantization methods. We use SmoothQuant~\cite{xiao2023smoothquant} as the basic weight-activation quantization method, and also compare with weight-only quantization algorithms including GPTQ~\cite{frantar2022gptq}, AWQ~\cite{lin2023awq} and Omniquant~\cite{shao2023omniquant}. Furthermore, we aggressively extend Omniquant to W4A4 and assess the corresponding accuracy degradation. Furthermore, we compare FMPQ with a recently released W4A8 KV4 quantization algorithm, QoQ~\cite{lin2024qserve},
which includes a group-wise quantization strategy (i.e., the group size is 128 and each group has one FP16 scale factor).

\textbf{Perplexity Evaluation. } Table \ref{tab:ppl_results} presents the evaluation results of Wikitext2 perplexity between FMPQ and other algorithm baselines. As one can notice, compared to W8A8 SmoothQuant and W4A16 Omniquant, FMPQ only introduces a slight perplexity increase (only 0.10 for LLaMA-1-7B). When we further introduce KV cache quantization, the increased perplexity is as small as 0.05 on average, which is negligible. According to our evaluation, only $16\%$ of the whole activations are quantized to 8-bit, on average. For LLaMA-30B, only $8\%$ of the total activations are quantized to 8-bit. Obviously, FMPQ provides a practical path for mixed-precision LLM serving. In comparison, when adopting a fully W4A4 Omniquant, the increased perplexity is unbearable (perplexity increase more than $5.21$), hindering the practical deployment of quantized LLMs.  

\textbf{Zero-shot Accuracy. } We further report the zero-shot accuracy of five common sense tasks in Table \ref{tab:zero_shot_results}. Compared with the state-of-the-art W4A16 quantization method, the accuracy acquired by FMPQ is only decreased by $0.75\%$. For LLaMA-3-8B, our FMPQ strategy even outperforms Omniquant when evaluating HellaSwag, demonstrating its efficiency. For LLaMA-3-8B, our fine-grained mixed-precision quantization strategy even outperforms the existing W4A8 KV4 quantization method, QoQ, demonstrating its efficiency. In summary, the FMPQ algorithm offers a viable path toward practical high-performance LLM serving.

\subsection{Kernel Performance}\label{sec:kernel-evaluation}

We evaluate the COMET-W4Ax kernel on various GEMM workloads within LLaMA families, Mistral-7B, Qwen2-72B and OPT-175B. We set three different batch sizes (16, 64 and 256) to validate the adaptability of different kernels. These various batch size configurations encompass a wide range of use cases, spanning from terminal devices to cloud data centers. During our kernel performance evaluation, the W4A4 ratio is set to $75\%$. Note that in actual model inference, we often achieve a higher W4A4 ratio. The baselines we compare include cuBLAS-W16A16~\cite{cublas}, TensorRT-W4A16 and TensorRT-W8A8~\cite{tensorrt-llm}. 

\begin{figure*}
    \centering
    \includegraphics[width=0.98\linewidth]{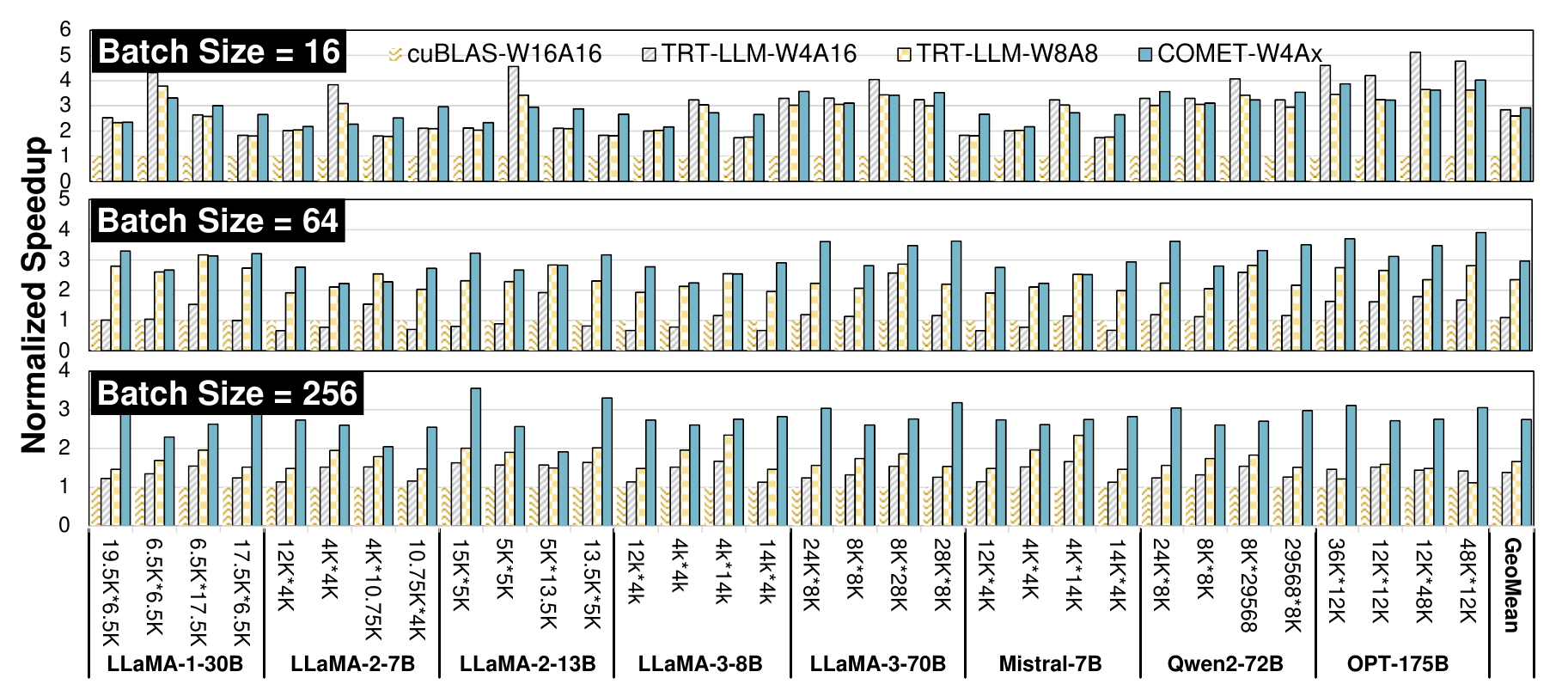}
    \caption{Linear layer speedups compared to the baselines for token generation phase.}
    \label{fig:kernel-performance}
\end{figure*}

\textbf{Results. }
Figure ~\ref{fig:kernel-performance} presents the latency speedups of COMET-W4Ax and other baselines. We use the performance of cuBLAS-W16A16 to normalize the performance of all GPU kernels. As one can notice, COMET-W4Ax achieves the best performance in most cases, particularly for large-batch parallelism (batch size = 256) processing cases. On average, COMET-W4Ax outperforms cuBLAS-W16A16, TRT-LLM-W4A116 and TRT-LLM-W8A8 by $2.88 \times, 1.77 \times$ and $1.33 \times$, respectively. TRT-LLM-W4A16 shows attractive performance with a small batch size (batch size = 16), which indicates that running LLM with small batch sizes is memory-bound. However, when the batch size is increased to 64 and 256, the performance gained by TRT-LLM-W4A16 is limited (only $1.10 \times$ and $1.38 \times$). In contrast, TRT-LLM-W8A8 exhibits poor performance with small batch sizes, but its speedup improves significantly as the batch size increases. Unlike these two kernels with varying performance, our proposed COMET-W4Ax achieves substantial gains across different batch sizes of 16, 64, and 256. Specifically, COMET-W4Ax achieves speedup factors of $2.91 \times, 2.97 \times$, and $2.75 \times$ for batch sizes of 16, 64, and 256, respectively.

\subsection{Ablation Study}\label{sec:ablation-study}

\begin{figure}
    \centering
    \includegraphics[width=0.98\linewidth]{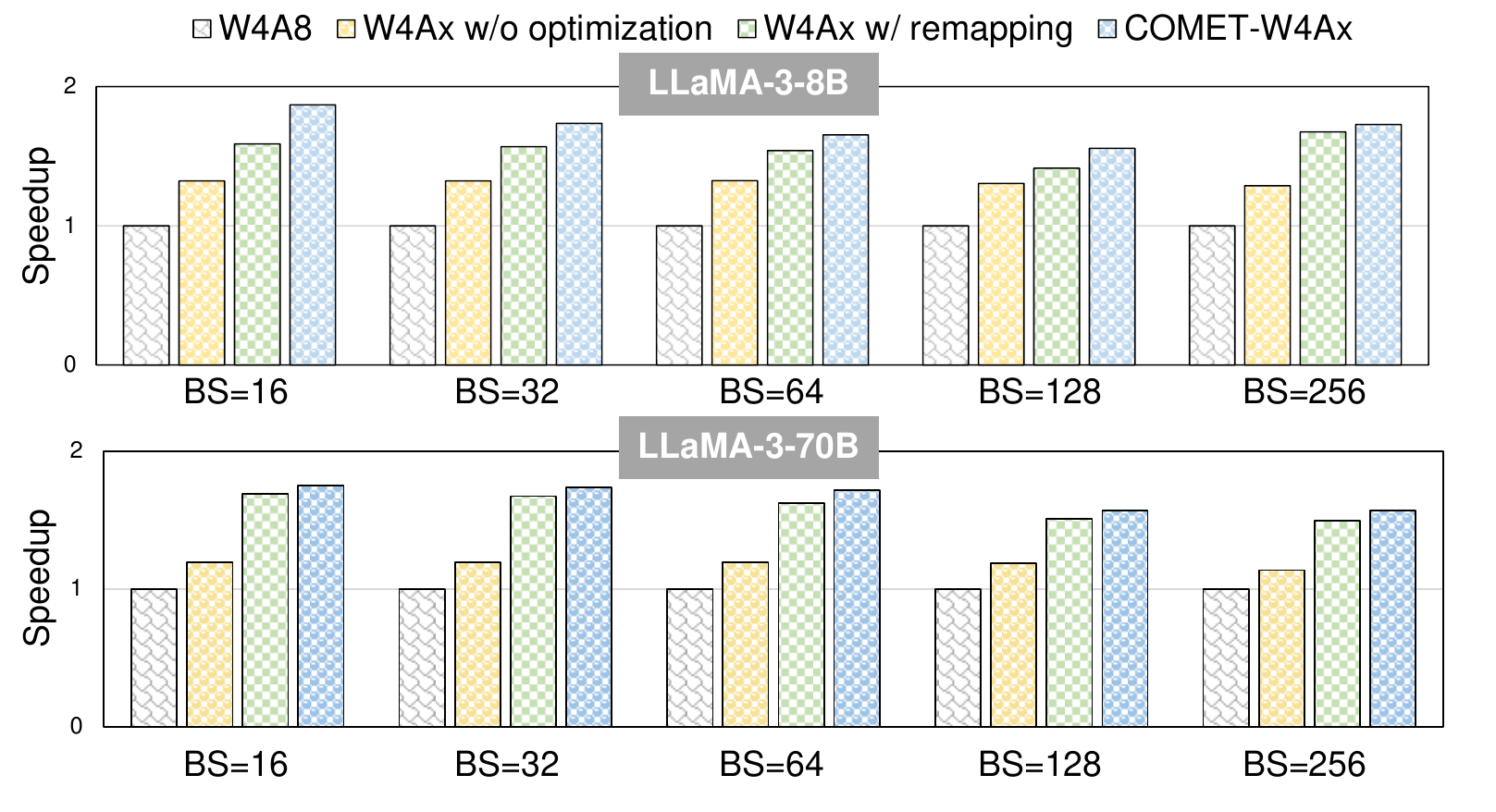}
    \caption{Explore the performance gains from different optimization strategies. A na\"ive implementation of W4Ax kernel can only achieve a limited performance improvement, while a highly optimized kernel can achieve a $1.69 \times$ speedup. }
    \label{fig:ablation-study}
\end{figure}

\begin{figure*}
    \centering
    \includegraphics[width=0.8\linewidth]{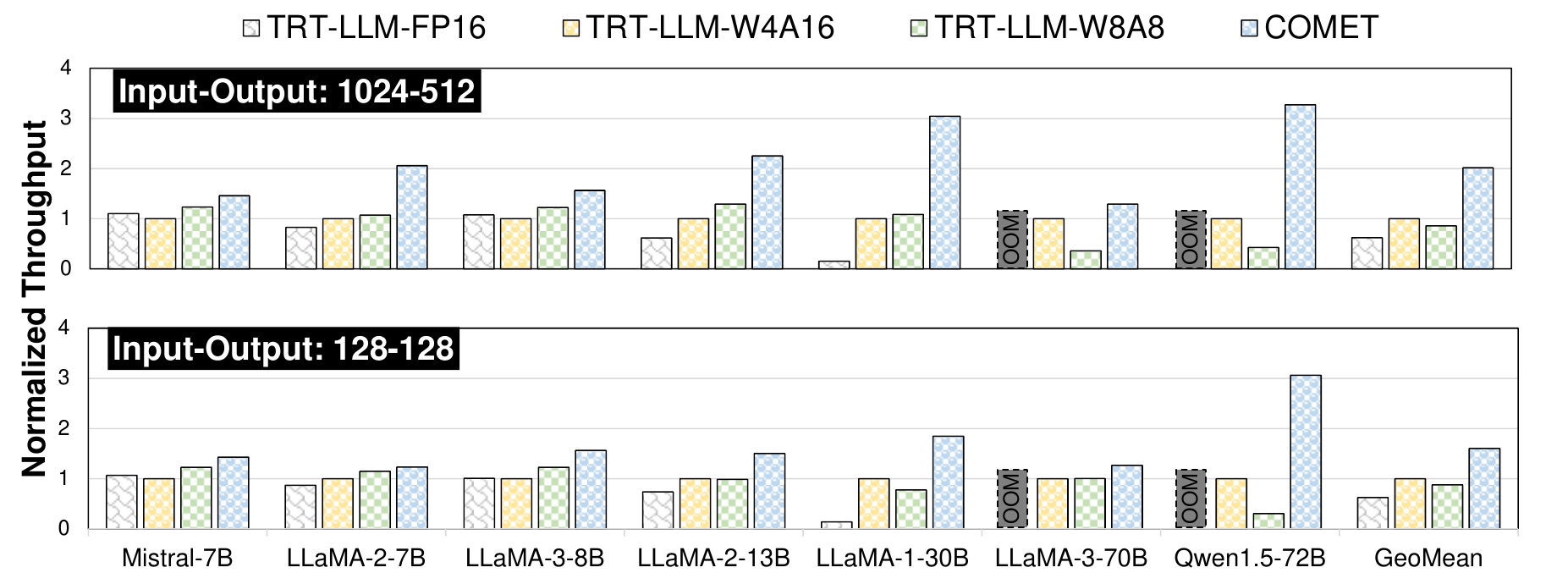}
    \caption{Compared with the TRT-LLM inference system, COMET provides higher throughput across various LLMs, ranging from 7B to 72B.}
    \label{fig:end2end-performance}
\end{figure*}

To support mixed-precision computation, we have developed a novel W4Ax kernel and employed various strategies to optimize the proposed COMET-W4Ax kernel, as detailed in Section \ref{sec:COMET-kernel}. To validate the effectiveness of these optimization strategies, we conducted an ablation study of the kernel on the LLaMA-3 family models across different batch sizes (from 16 to 256). We evaluate several versions of the kernel, including W4A8, a na\"ive W4Ax kernel without any optimizations (W4Ax w/o optimization), a kernel implementing the tile remapping (W4Ax w/ remapping) strategy proposed in Section \ref{sec:sm-scheduling}, and a fully optimized COMET-W4Ax kernel.

Figure \ref{fig:ablation-study} illustrates the performance differences between the COMET-W4Ax and the under-optimized W4Ax kernel. Compared to the W4A8 GEMM kernel, a na\"ive implementation of the W4Ax kernel achieves only $1.31 \times$ and $1.18 \times$ speedups. These potential gains stem from the use of INT4 tensor cores, which offer $2 \times$ higher throughput. However, due to the lack of load balancing across different SMs, the utilization of INT4 tensor cores falls short of expectations. By implementing tile and SM remapping, the speedup increases to $1.56 \times$ and $1.60 \times$, respectively. By further eliminating the one-to-one binding between tiles and SMs, COMET-W4Ax achieves $1.71 \times$ and $1.67 \times$ speedups for GEMM computations on the LLaMA-3-8B and LLaMA-3-70B models, respectively. Furthermore, the proposed COMET-W4Ax kernel achieves significant performance improvements across various batch sizes, demonstrating its broad applicability.

\subsection{End-to-End Evaluation}\label{sec:end-to-end}

We explore the maximum achievable throughput of different inference systems, within the same memory constraints on a single A100-80G-SXM4. Specifically, we adopt two different settings, including an input/output sequence length of 1024/512 and an input/output sequence length of 128/128, to evaluate models including Mistral-7B, LLaMA family models and Qwen1.5-72B. Furthermore, we also compare the normalized throughput under the same batch sizes for LLaMA-3-8B.

\textbf{Throughput Evaluation.} Figure \ref{fig:end2end-performance} presents the relative throughput performance of different inference systems. We set the TRT-LLM-W4A16 as the baseline. According to our evaluation, COMET achieves $2.02 \times$ and $1.63 \times$ higher throughput on average for two different input/output sequence length settings, respectively. With the help of low-precision quantization on weight, activation and KV cache, we can easily support large-batch parallelism for large models such as LLaMA-3-70B and Qwen1.5-72B. Specifically, relative to the best-performing configurations (either W4A16 or W8A8), COMET demonstrates impressive performance improvement: it achieves $1.18 - 1.93 \times$ higher throughput for 7B and 8B models, $1.74 \times$ higher throughput for LLaMA-2-13B, $2.81 \times$ higher throughput for LLaMA-1-30B and $1.28 - 3.27 \times$ higher throughput for 70B and 72B models. COMET performs better when processing tasks with longer output sequences (512 tokens), as our proposed 4-bit KV cache can effectively migrate the bottleneck of large batch execution. Even for smaller models (e.g., 8B) and shorter output sequence lengths (input/output of 128/128), COMET still delivers substantial performance gains. For example, compared to TRT-LLM-W4A16, COMET achieves a $1.56 \times$ throughput improvement on the LLaMA-3-8B model. This enhancement is primarily due to the efficient utilization of the INT4 tensor cores in the A100, providing higher computational efficiency.

\begin{figure}
    \centering
    \includegraphics[width=0.98\linewidth]{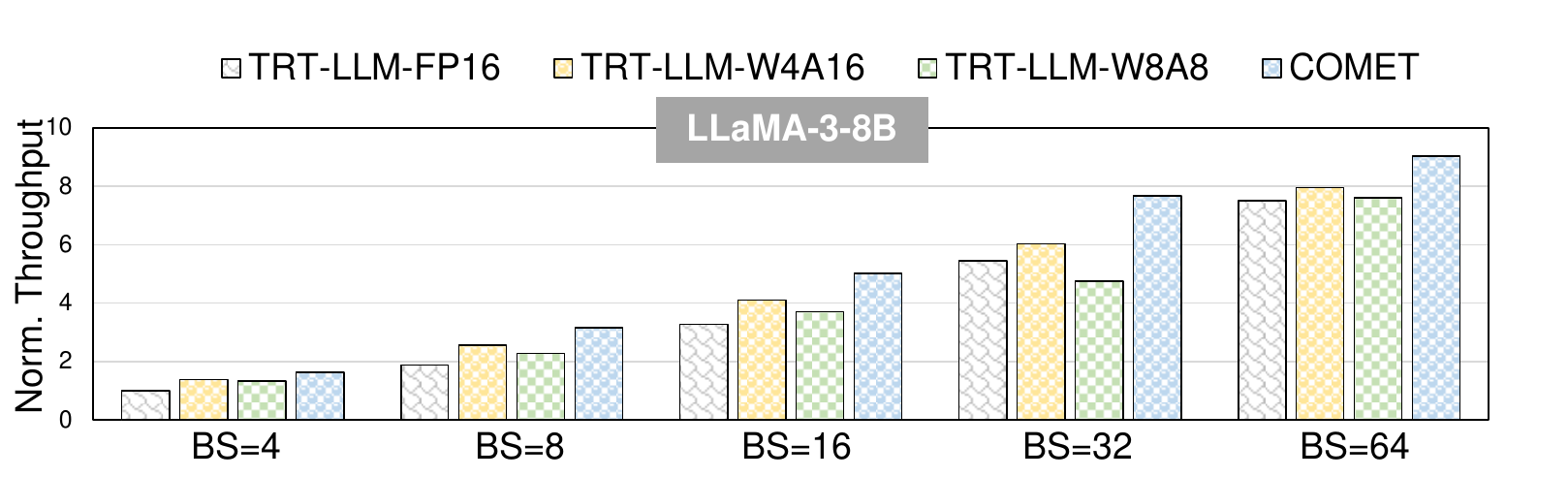}
    \caption{Same-batch throughput comparison between COMET and baseline inference system for LLaMA-3-8B. We use an input sequence length of 1024 and output sequence length of 512.}
    \label{fig:same-batch-throughput}
\end{figure}

\textbf{Comparisons under the same batch sizes. } We demonstrate the speedup results under the same batch sizes for LLaMA-3-8B in Figure \ref{fig:same-batch-throughput}. As the batch size increases, the execution throughput gradually improves. For example, when setting the batch size as 64 for TRT-LLM-FP16, the throughput is increased by $7.52 \times$ compared with the batch size of 4. Hence, supporting large-batch parallelism is essential for modern GPUs. Following the setting in COMET, we can achieve even larger batch sizes. Furthermore, under the same batch sizes, COMET consistently outperforms the best configurations of TensorRT-LLM. According to our evaluation, COMET achieves a $1.37 \times$ speedup than SOTA TensorRT-LLM configurations. This is primarily due to our efficient utilization of INT4 tensor cores and fast dequantization.



\section{Conclusion}\label{sec:conclusion}
In this paper, we present COMET, the first mixed-precision LLM inference framework designed for practical low-bit mixed-precision LLM serving, primarily built on the proposed FMPQ algorithm and the COMET-W4Ax kernel. Specifically, the FMPQ algorithm efficiently achieves low-precision quantization for activations and the KV cache with minimal accuracy loss. Moreover, the open-source COMET-W4Ax kernel can be seamlessly integrated into existing inference systems. It includes optimizations for data layout, GPU software pipeline, and streaming multiprocessor scheduling, addressing data access and load imbalance issues in mixed-precision GEMM on modern GPUs. Evaluations on a single A100-80G-SXM4 demonstrate that COMET achieves a $2.02\times$ end-to-end throughput improvement over state-of-the-art baselines, showcasing its potential for enhancing LLM inference efficiency.


\bibliographystyle{plain}
\bibliography{references}

\end{document}